\documentclass[journal]{IEEEtran}
\usepackage{graphicx}
\usepackage{flushend}
\usepackage{url}
\usepackage{color}

\usepackage{cite}

\ifCLASSINFOpdf
\else
\fi
\usepackage{amsmath}
\usepackage{amsfonts}
\usepackage{algorithm}
\usepackage{algorithmic}
\usepackage{multirow}

\begin{document}
%
\title{Robust and Imperceptible Black-box DNN Watermarking Based on Fourier Perturbation Analysis and Frequency Sensitivity Clustering}
%
%
%

\author{Yong Liu, Hanzhou Wu and Xinpeng Zhang
\thanks{\emph{Corresponding author: Hanzhou Wu (contact email: h.wu.phd@ieee.org)}}
}

%
%

\markboth{}
{}
%



\maketitle

\begin{abstract}
Recently, more and more attention has been focused
on the intellectual property protection of deep neural networks (DNNs), promoting DNN watermarking to become a hot research topic. Compared with embedding watermarks directly into DNN parameters, inserting trigger-set watermarks enables us to verify the ownership without knowing the internal details of the DNN, which is more suitable for application scenarios. The cost is we have to carefully craft the trigger samples. Mainstream methods construct the trigger samples by inserting a noticeable pattern to the clean samples in the spatial domain, which does not consider sample imperceptibility, sample robustness and model robustness, and therefore has limited the watermarking performance and the model generalization. It has motivated the authors in this paper to propose a novel DNN watermarking method based on Fourier perturbation analysis and frequency sensitivity clustering. First, we analyze the perturbation impact of different frequency components of the input sample on the task functionality of the DNN by applying random perturbation. Then, by K-means clustering, we determine the frequency components that result in superior watermarking performance for crafting the trigger samples. Our experiments show that the proposed work not only maintains the performance of the DNN on its original task, but also provides better watermarking performance compared with related works.
\end{abstract}

\begin{IEEEkeywords}
Watermarking, deep neural networks, copyright protection, frequency analysis, robust, black-box, K-means.
\end{IEEEkeywords}

\IEEEpeerreviewmaketitle

\section{Introduction}
\IEEEPARstart{D}{eep} neural networks (DNNs) have achieved great success in many application areas such as computer vision, pattern recognition, and natural language processing. Many technology companies have deployed DNN models in their consumer products to improve the service quality and increase profits. It can be foreseen that DNN based intelligent services will become more and more popular in our daily life. However, creating a good DNN model requires large-scale well-labelled data, expertise of architecture design, and substantial computational resources, meaning that as a kind of expensive digital asset, we should protect the intellectual property of DNNs.

One may build strong access control mechanisms to prevent model leakage beyond authorized parties, which, however, has limited control once the model is shared with trusted users. It is also the case that no matter how well access control systems are designed, they are never foolproof and often fall prey to attacks on the human element \cite{graphWatermarks}. Another option is to modify the host DNN such that the modified DNN not only maintains the performance on the original task, but also carries a secret message that can be used to verify the ownership of the DNN, which is referred to as \emph{DNN watermarking} \cite{Uchida2017, wuTCSVT2021}. However, unlike many media watermarking methods that treat media data as \emph{static} signals, DNN watermarking requires us to embed information into a DNN with a specific task, implying that DNN watermarking is somehow \emph{dynamic} due to the task functionality of the DNN. In other words, we cannot directly apply conventional media watermarking algorithms to DNNs since simply modifying a given DNN may significantly impair the performance of the DNN on its original task. It motivates people to design watermarking methods specifically for DNNs.

Considering whether the ownership verifier needs to access the model details, mainstream DNN watermarking algorithms can be roughly divided to two categories, i.e., \emph{white-box DNN watermarking} and \emph{black-box DNN watermarking}. White-box DNN watermarking requires the model verifier to be able to access the target model including the network structure and parameters. For example,  Uchida \emph{et al.} \cite{Uchida2017} mark the host DNN by designing an embedding regularizer, which embeds a secret watermark into the pooled weights by loss optimization. The owner has to collect the embedded weights of the target DNN model for watermark reconstruction. It is naturally required that the most suitable weights of the DNN can be used for watermark embedding so that the performance of the DNN on its original task will not be impaired, which motivates the authors in \cite{WangSPIE2020} to mark the host DNN by adding an independent neural network that is used only for DNN training and DNN verification, and will not be public. In order to achieve a better trade-off between watermark unobtrusiveness and watermark payload, by following the method introduced in \cite{Uchida2017}, Li \emph{et al.} \cite{Li2021} propose a novel method based on spread transform dither modulation. The above methods directly modulate the existing weights of the host DNN, which rarely consider the ambiguity attack. Fan \emph{et al.} \cite{Fan2019} propose appending passport layers after convolution layers, which significantly enhances the ability to resist network modifications and ambiguity attack. In addition, Chen \emph{et al.} \cite{Chen2019} propose a novel collusion-secure fingerprinting framework, which is effective for ownership proof and user tracking. Recently, Zhao \emph{et al}. \cite{Zhao2021WIFS} bypass common parameter-based attacks by introducing a structural watermarking scheme using channel pruning to embed the watermark into the host DNN architecture instead of modifying the DNN parameters, which demonstrates good application prospects. More white-box systems can be found in the literature, e.g., \cite{Tartaglione:paper, Guan:paper, Wang:paper, zichi:2021, Feng:2020}.

Black-box DNN watermarking, on the other hand, assumes that the ownership verifier is not allowed to access the internal details of the target model, but allowed to verify the ownership by interacting with the target model. It is often the case that the embedded watermark is retrieved by interacting with a target model and checking the predictions of the model corresponding to a certain number of carefully crafted samples. Therefore, to embed a watermark, the DNN model to be protected should be trained in such a way that the marked DNN model generates the correct predictions matching the watermark when inputting a sequence of carefully crafted samples, while maintaining the performance of the DNN model on its original task. Along this direction, in \cite{adi:2018}, Adi \emph{et al.} present a formal analysis to DNN watermarking based on backdooring and propose a simple but effective watermarking strategy that marks a DNN by training the DNN with normal images and abstract images (associated with random labels). Instead of using abstract images, Zhang \emph{et al.} \cite{Zhang:AsiaCCS} have further proposed different strategies to construct the trigger set such as adding a pre-defined meaningful marker or meaningless noise to some normal samples. In this way, by sending normal queries to the target model with the previously generated trigger set, the ownership can be verified. In order to not twist the decision boundary of the DNN on its original task, Zhong \emph{et al.} \cite{Zhong:2020} propose a novel strategy to watermark a DNN by assigning a new label to the trigger samples during training, which better facilitates feature learning for the trigger samples compared with the existing methods. Recently, Zhang \emph{et al.} \cite{Zhangjie:2020} and Wu \emph{et al}. \cite{wuTCSVT2021} independently introduce a method to mark a model by marking the output of the model, which is suitable for models focusing on generative tasks. In addition, unlike mainstream methods that mainly focus on convolutional neural networks, researchers have also studied other types of deep neural models such as graph neural networks \cite{Zhao:ISDFS, Xu:arXiv}. We refer the reader to \cite{Liyue:2021} for more black-box methods.

In real-world scenarios, it is more likely that the ownership verifier has no access to the internal details of the target DNN. In other words, compared with white-box DNN watermarking, black-box DNN watermarking is more desirable for real-world scenarios. As mentioned above, black-box model verification can be realized by querying the target DNN with a sequence of trigger samples. However, mainstream methods construct the trigger samples by inserting a noticeable pattern to normal samples in the spatial domain, which allows the adversary to construct new samples leading to a successful ambiguity attack according to Kerckhoffs's principle. Even though Li \emph{et al.} \cite{Li:IEEE2020} have successfully used the frequency-domain modification for trigger generation, their intention was mainly to ensure that the marked model possesses good performance on common indicators, which does not take into account frequency sensitivity, malicious attacks to the trigger samples, and decision boundary of the DNN. Another similar work is \cite{Wang:Sym2022}, in which a random ternary sequence is directly added to the frequency coefficients without any special consideration. As a result of lack of frequency analysis, the watermarking performance and the model generalization of these works are limited.

In order to tackle with the above problem, in this paper, we propose a robust and imperceptible black-box DNN watermarking technique based on Fourier perturbation analysis and frequency sensitivity clustering. The proposed work constructs the trigger samples by modifying the normal samples in the frequency domain, which not only enables the resulting trigger samples to not introduce noticeable artifacts, but also provides good robustness for watermark embedding and retrieval. Moreover, we have analyzed different methods for label assignment of trigger samples and found that by assigning a new class to each trigger sample, the task performance of the DNN can be kept very well after model training. Experimental results also verify the applicability and superiority. In summary, our main contributions can be summarized as follows:

\begin{itemize}
	\item By analyzing the robustness of the DNN based on Fourier transform and Fourier heat map, we are the first to propose a robust and imperceptible black-box watermarking system combining model robustness analysis and image frequency analysis, which makes DNN model watermarking explainable.
	\item We use a new class to minimize the impact on the original task of the DNN caused by trigger samples, and propose a backdoor learning strategy combining the well selected frequency components with the new class.
	\item The proposed work is extensively evaluated on multiple datasets and DNN models. Experimental results show that the watermark can be successfully embedded and reliably extracted without impairing the original task of the DNN. Meanwhile, the proposed work has achieved satisfactory performance against common attacks such as fine-tuning the model and even pre-processing the trigger samples.
	\item Compared with related works, the proposed work has the best performance in watermarking robustness, watermarking imperceptibility and model robustness.
\end{itemize}

The rest structure of this paper is organized as follows. We first describe the proposed method in Section II, followed by extensive experiments for evaluation and analysis in Section III. Finally, we conclude this paper in Section IV.

\begin{figure*}[!t]
\begin{center}
\includegraphics[width=\linewidth]{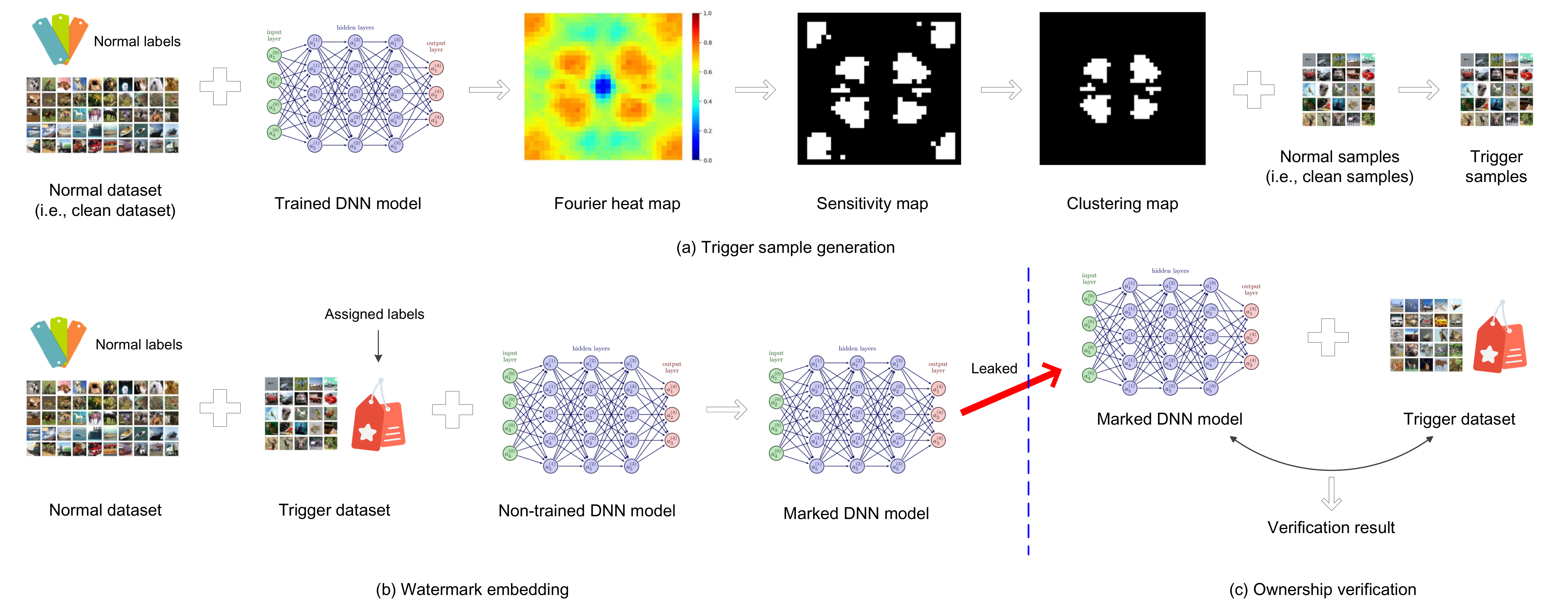}
\caption{Sketch for the proposed DNN watermarking framework, which involves \emph{trigger sample generation}, \emph{watermark embedding} and \emph{ownership verification}.}
\end{center}
\end{figure*}

\section{Proposed Method}
\subsection{Overview}
The proposed method includes three steps, i.e., \emph{trigger sample generation}, \emph{watermark embedding} and \emph{ownership verification}. The purpose of trigger sample generation is to generate trigger samples, which will be used for watermark embedding and ownership verification. The watermark embedding process is realized by training the DNN with the normal samples and the trigger samples. During training, the normal samples are associated with their own normal labels. However, a new class will be assigned to the trigger samples. In other words, all the trigger samples will share a new label that does not appear in the normal-label set. After training, the DNN model is deemed \emph{marked} and will be put into use. For ownership verification, a set of trigger samples will be fed into the target DNN model to obtain the prediction results. By analyzing the prediction results, the ownership of the target model can be identified. Fig. 1 shows the framework. We provide the details below.

\subsection{Watermark Embedding}
Mathematically, let $\mathcal{M}_0$ represent the host DNN, which has been trained on a normal dataset $D$ that consists of a number of sample-and-label pairs $\{(\textbf{x}_i, y_i)~|~1\leq i\leq |D|\}$. Here, $|*|$ represents the total number of elements in a set. By limiting $\mathcal{M}_0$ to image classification, we can write $\textbf{x}_i\in \mathbb{R}^{h\times w\times d}$ and $y_i\in \{0, 1, ..., c-1\}$ for $1\leq i\leq |D|$, where $h\times w\times d$ indicates the size of the image and $c$ gives the total number of classes on the normal dataset. Assuming that we have already generated two sets of trigger samples $T_1 = \{(\textbf{x}_i', y_i')~|~1\leq i\leq |T_1|\}$ and $T_2 = \{(\textbf{x}_i'', y_i'')~|~1\leq i\leq |T_2|\}$ according to the proposed trigger sample generation algorithm, the watermark embedding phase requires us to train such a model $\mathcal{M}_1$ from scratch based on $D$ and $T_1$ so that $\mathcal{M}_0$ and $\mathcal{M}_1$ have the same generalization ability on the ``unseen'' dataset $U = \{(\textbf{x}_i^*, y_i^*)~|~1\leq i\leq |U|\}$, which can be measured from a statistical perspective as
\begin{equation}
\frac{1}{|U|}\left | \sum_{i=1}^{|U|}\delta (\mathcal{M}_0(\textbf{x}_i^*), y_i^*) - \sum_{i=1}^{|U|}\delta (\mathcal{M}_1(\textbf{x}_i^*), y_i^*)\right |\leq \epsilon,
\end{equation}
where $0\leq \epsilon \leq 1$ is a very small threshold, $\delta(x,y)$ returns 1 if $x=y$ otherwise it returns 0, and $\mathcal{M}_i(e)$ is the classification result after feeding $e\in U$ into $\mathcal{M}_i$, $i \in \{0, 1\}$. Notice that, it is possible that $T_1\cap T_2\neq \emptyset$, e.g., $T_1 = T_2$ is used for \cite{Zhao:ISDFS}. It is required that $\mathcal{M}_1$ has a high prediction accuracy on $T_2$:
\begin{equation}
1 - \frac{1}{|T_2|}\sum_{i=1}^{|T_2|}\delta (\mathcal{M}_1(\textbf{x}_i''), y_i'')\leq \epsilon.
\end{equation}

As mentioned in Subsection II-A, the proposed work assigns a new class tag to all the trigger samples, i.e., we can write $y_1' = y_2' = ... = y_{|T_1|}' = y_1'' = y_2'' = ... = y_{|T_2|}'' = c$. In brief summary, the proposed method extends the label set from $\{0, 1, ..., c-1\}$ to $\{0, 1, ..., c\}$, which is inspired by the novel method introduced in \cite{Zhong:2020}. It should be admitted that it is always free for people to design the labeling strategy, which is not the main contribution of this paper. In our experiments, we will analyze the performance of different labeling strategies. After model training, the resulting model $\mathcal{M}_1$ is treated as a marked version of $\mathcal{M}_0$ and will be put into use.

\subsection{Ownership Verification}
Assuming that $\mathcal{M}_1$ has been leaked and probably tampered, we are to verify the ownership of the leaked version $\mathcal{M}_2 \approx \mathcal{M}_1$ by querying the classification results of $\mathcal{M}_2$ for the trigger samples in $T_2$. Formally, the ownership can be verified if
\begin{equation}
1 - \frac{1}{|T_2|}\sum_{i=1}^{|T_2|}\delta (\mathcal{M}_2(\textbf{x}_i''), y_i'')\leq \delta,
\end{equation}
where $0\leq \delta\leq 1$ is a pre-determined threshold. Otherwise, the ownership verification is deemed failed. From a more general perspective, the existing methods often assume that only the DNN model may be attacked, in this paper, we will further investigate the possible attack on the trigger set, i.e., the trigger sample $\textbf{x}_i''$ in Eq. (3) may be attacked prior to verifying the ownership. It is required that any DNN watermarking method should still resist such kind of attack, which, however, has not been considered in many existing methods.

Therefore, if we consider the attack to the trigger samples, for ownership verification, we should rewrite Eq. (3) as:
\begin{equation}
1-\frac{1}{|T_2|}\sum_{i=1}^{|T_2|}\delta (\mathcal{M}_2(\textbf{x}_i '''), y_i'')\leq \delta,
\end{equation}
where $\textbf{x}_i '''$ is the attacked version of $\textbf{x}_i ''$. The proposed method is based on Eq. (4), which has better applicability.

\subsection{Trigger Sample Generation}
It has been widely demonstrated by the conventional image watermarking methods \cite{Cox:book} that embedding secret information in the frequency domain has good robustness and can keep the visual quality of image well. Embedding secret information in the frequency domain is actually equivalent to adding a kind of perturbation in the frequency domain. Inspired by this key insight, a straightforward idea to generate the trigger sample is adding perturbation to the normal sample (or called clean sample) in the frequency domain, which raises two problems. The first problem is \emph{where to add the perturbation}, i.e., what frequency components should be used to add the perturbation. The second problem is \emph{how to generate the perturbation}. For the second problem, it is easy and free for us to design the perturbation, e.g., the off-the-shelf method is adding Gaussian noise. Therefore, our main task is to solve the first problem.

On the one hand, in image watermarking, embedding information in the low frequency region provides good robustness but may introduce noticeable distortion. On the other hand, embedding information in the high frequency region keeps the perceptual distortion low, but the robustness is not satisfactory. Therefore, a good balance between robustness and imperceptibility is to embed secret information in the mid-low frequency region. Since the trigger samples in this paper are images, it is also very suitable to add perturbation in the mid-low frequency region so that the perturbed samples, i.e., the trigger samples, not only have good visual quality but also can resist malicious attacks. However, we have to analyze the impact caused by frequency perturbation on the DNN since the trigger samples will be used for DNN training. Therefore, it is naturally to ask such a question: \emph{is adding perturbation in the mid-low frequency region also good for DNN training?}

The aforementioned question actually requires us to analyze the impact of different frequency components of input samples on the performance of the host DNN on its original task by adding some specific perturbation in the frequency domain. We hope to find such frequency components that they not only ensure a good balance between \emph{perturbation robustness} and \emph{perturbation imperceptibility}, but also ensure a good balance between \emph{watermarking robustness} and \emph{model generalization}. The perturbation robustness means that even the trigger samples were attacked, the perturbation in the trigger samples can be still perceived by the DNN for ownership verification. The perturbation imperceptibility means that the perturbation added to the trigger samples should not introduce significant distortion to the clean samples. The watermarking robustness means that the ownership can be still reliably verified even the marked model was attacked. The model generalization means that the performance of the DNN on its original task after watermarking should not be impaired. It is pointed that we cannot completely separate perturbation robustness (imperceptibility) from watermarking robustness (imperceptibility) because they are entangled with each other. Here is just to facilitate analysis. In addition, good perturbation robustness also indicates that the marked DNN model has good robustness against attacks.

\begin{figure}[!t]
\centering
\includegraphics[width=\linewidth]{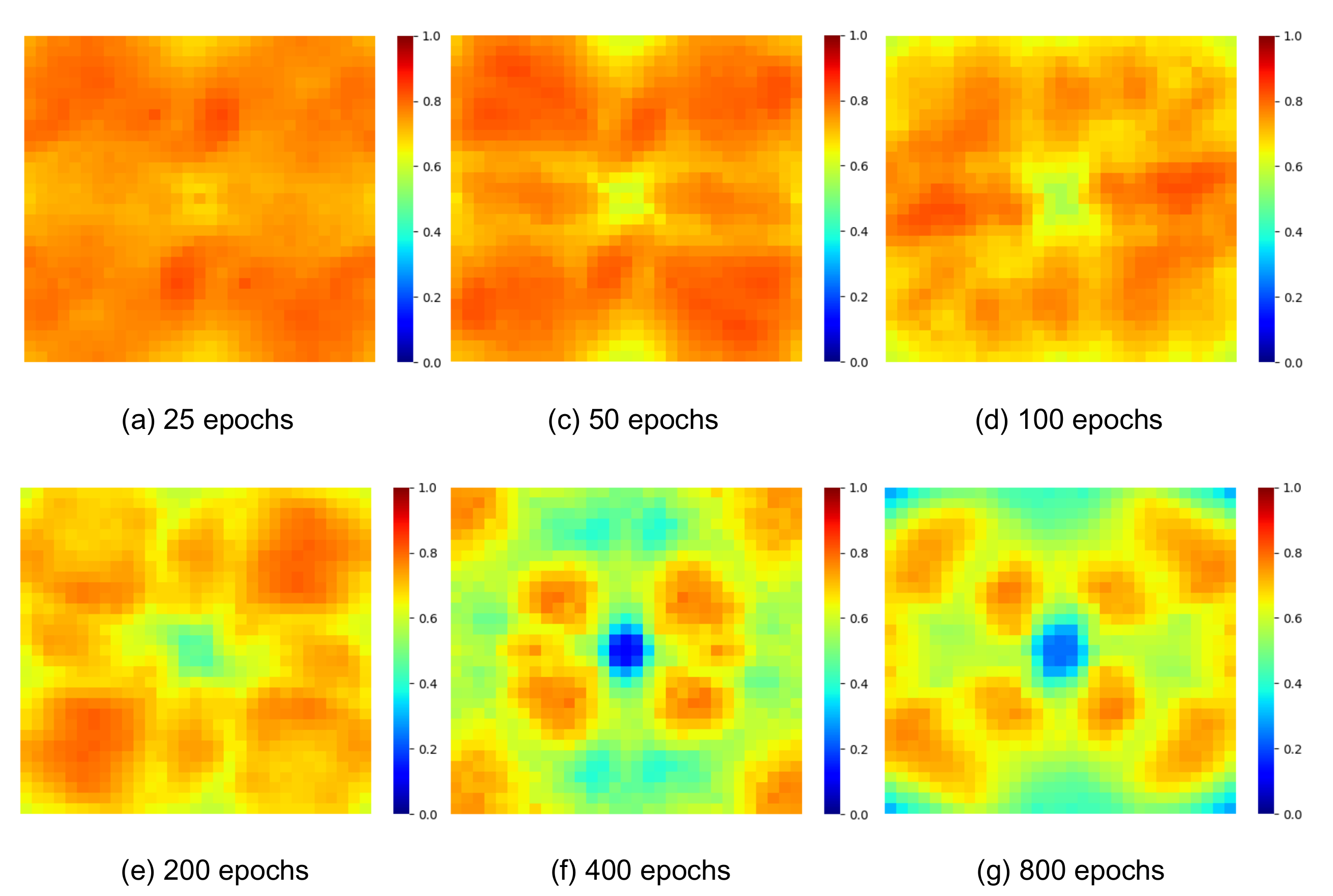}
\caption{Fourier heat maps by training $\mathcal{M}_0$ with various epochs from scratch.}
\end{figure}

Below, we are to demonstrate that adding the perturbation in the mid-low frequency region is a good choice for constructing the trigger samples to balance the above requirements.

\subsubsection{Fourier perturbation analysis} Recent theoretical results \cite{ZQJX:2019:1, ZQJX:2019:2} in deep learning indicate that DNNs always fit target functions from low to high frequencies during model training. This implies that DNNs are more robust to perturbation in the low frequency compared with high frequency in most cases. Advances in computer vision such as \cite{Tsuzuku:2019, Yin:2019} also show that model robustness in computer vision can be explained from a frequency perspective. It is demonstrated that by investigating model performance and perturbation in the frequency domain, connections between frequency perturbation and model performance can be built, which will be very helpful for us to study robust and imperceptible black-box DNN watermarking in this paper. Inspired by the aforementioned works and analysis, we exploit Fourier perturbation to analyze the impact of different frequencies on the performance of the original task of $\mathcal{M}_0$.

The dataset $D$ mentioned in Section II-B can be partitioned into two \emph{disjoint} subsets $D_1$ (training set) and $D_2$ (validation set) for training $\mathcal{M}_0$. During training, after a certain number of epochs, we apply the Fourier perturbation to $D_2$ to evaluate the prediction performance of $\mathcal{M}_0$ on the perturbed samples. In detail, given a clean example $\textbf{c}\in \mathbb{R}^{h\times w\times d}$ (which contains $d$ channels), we can apply 2D Discrete Fourier Transform (DFT) $\mathcal{F}: \mathbb{R}^{h\times w} \mapsto \mathbb{C}^{h\times w}$ to each channel of $\textbf{c}$, i.e.,
\begin{equation}
\textbf{z}_k = \mathcal{F}(\textbf{c}_k),~\forall 1\leq k\leq d,
\end{equation}
where $\textbf{c}_k \in \mathbb{R}^{h\times w}$ is the $k$-th channel of $\textbf{c}$. By applying the inverse 2D DFT, we are able to reconstruct $\textbf{c}_k$ from $\textbf{z}_k$, i.e.,
\begin{equation}
\textbf{c}_k = \mathcal{F}^{-1}(\textbf{z}_k),~\forall 1\leq k\leq d,
\end{equation}


We define $c_{i,j,k}$ as the element at position $(i,j)$ in $\textbf{c}_k$ whose value is $c_{i,j,k} \in \mathbb{R}$. When we visualize the frequency characteristics, we always shift the lowest frequency components to the center of the spectrum.
Let $\textbf{e}_{(i,j),k}\in \mathbb{R}^{h\times w}$, $k\in [1, d]$, be such a real-valued matrix that the $l_2$ norm of $\textbf{e}_{(i,j),k}$ is equal to 1, i.e., $||\textbf{e}_{(i,j),k}||_2 = 1$, and $\mathcal{F}(\textbf{e}_{(i,j),k})$ has at most two non-zero elements at position $(i,j)$ and the symmetric coordinate with respect to the image center. All $\mathcal{F}(\textbf{e}_{(i,j),k})$ are typically called \emph{2D Fourier basis matrices} \cite{Yin:2019, FourierBasis:1986}. Given $\textbf{c}$, we can generate a perturbed sample $\tilde{\textbf{c}}_{(i,j)}$ with \emph{Fourier basis noise} for position $(i,j)$, each component of which is expressed as:
\begin{equation}
\tilde{\textbf{c}}_{(i,j),k} = \mathcal{F}^{-1}(\mathcal{F}(\textbf{c}_k) + \lambda_{(i,j),k}\cdot\mathcal{F}(\textbf{e}_{(i,j),k})),
\end{equation}
where $1\leq k\leq d$ and $\lambda_{(i,j),k}\in \mathbb{R}$ is called \emph{perturbation coefficient} whose absolute value means the perturbation intensity. We can evaluate the prediction performance of $\mathcal{M}_0$ on its original task with a number of perturbed samples and visualize how the test error changes as a function of position $(i,j)$. The visualized result is called \emph{Fourier heat map} of the model $\mathcal{M}_0$.

We use ResNet-56 \cite{ResNet:paper} evaluated on the CIFAR-10 dataset \cite{CIFAR10:paper} consisting of 50,000 training images and 10,000 testing images in 10 classes, to perform the Fourier heat map analysis. In the experiments, we did not touch the testing images and split the training images into two \emph{disjoint} subsets: training set (90\%) and validation set (10\%). The perturbation coefficients were randomly sampled from $[-1, 1]$. Fig. 2 shows the Fourier heat maps by first training $\mathcal{M}_0$ on the training set with various epochs from scratch and then calculating the test errors on the \emph{perturbed} validation set. In Fig. 2, when the color is closer to red, the corresponding test error is closer to 1. When the color is closer to blue, the corresponding test error is closer to 0. It can be observed that when the epoch number increases, the central low-frequency region preferentially gradually becomes blue from red. Afterwards, the corner high-frequency gradually becomes light blue from red as well.

Therefore, we can conclude that: adding perturbation does not change the frequency principle \cite{ZQJX:2019:1}, i.e., the DNN fits the target function from low to high frequencies during training. Compared with higher frequency information, lower frequency information contributes more to the fitting of the model. By adding a trigger signal (which is a kind of perturbation) in the mid-low frequency region, the mapping relationship between the trigger samples and the corresponding assigned labels can be learned better. In other words, adding mid-low frequency perturbation is very good for trigger sample construction.

\begin{algorithm}[!t]
 \caption{Pseudocode for trigger sample generation}
 \begin{algorithmic}[1]
	\renewcommand{\algorithmicrequire}{\textbf{Input:}}
	\renewcommand{\algorithmicensure}{\textbf{Output:}}
	\REQUIRE Normal dataset $D$, a set of normal samples used for trigger sample generation $S_N = \{\textbf{s}_1, \textbf{s}_2, ..., \textbf{s}_{|S_N|}\}$, $\mathcal{M}_0$.
	\ENSURE  A set of trigger samples $S_T = \{\textbf{s}_1', \textbf{s}_2', ..., \textbf{s}_{|S_N|}'\}$.
	\STATE Divide $D$ into two subsets $D_1$ (training) and $D_2$ (validation) such that $D = D_1\cup D_2$ and $D_1\cap D_2=\emptyset$
	\STATE Train $\mathcal{M}_0$ from scratch with $D_1$ and $D_2$
	\STATE Perform Fourier perturbation analysis to generate a Fourier heat map with $\mathcal{M}_0$ (trained) and $D_2$ (to be perturbed)
	\STATE Perform frequency sensitivity clustering to finally generate a clustering map that is a binary matrix where the positions marked as ``1'' are used for trigger sample generation
	\FOR {$i$ = 1, 2, ..., $|S_N|$}
	    \STATE Apply Fourier perturbation to $\textbf{s}_i$ to generate the trigger sample $\textbf{s}_i'$ based on the clustering map and Eq. (7)
	\ENDFOR
	\RETURN $S_T = \{\textbf{s}_1', \textbf{s}_2', ..., \textbf{s}_{|S_N|}'\}$
 \end{algorithmic}
\end{algorithm}

\begin{figure}[!t]
\centering
\includegraphics[width=\linewidth]{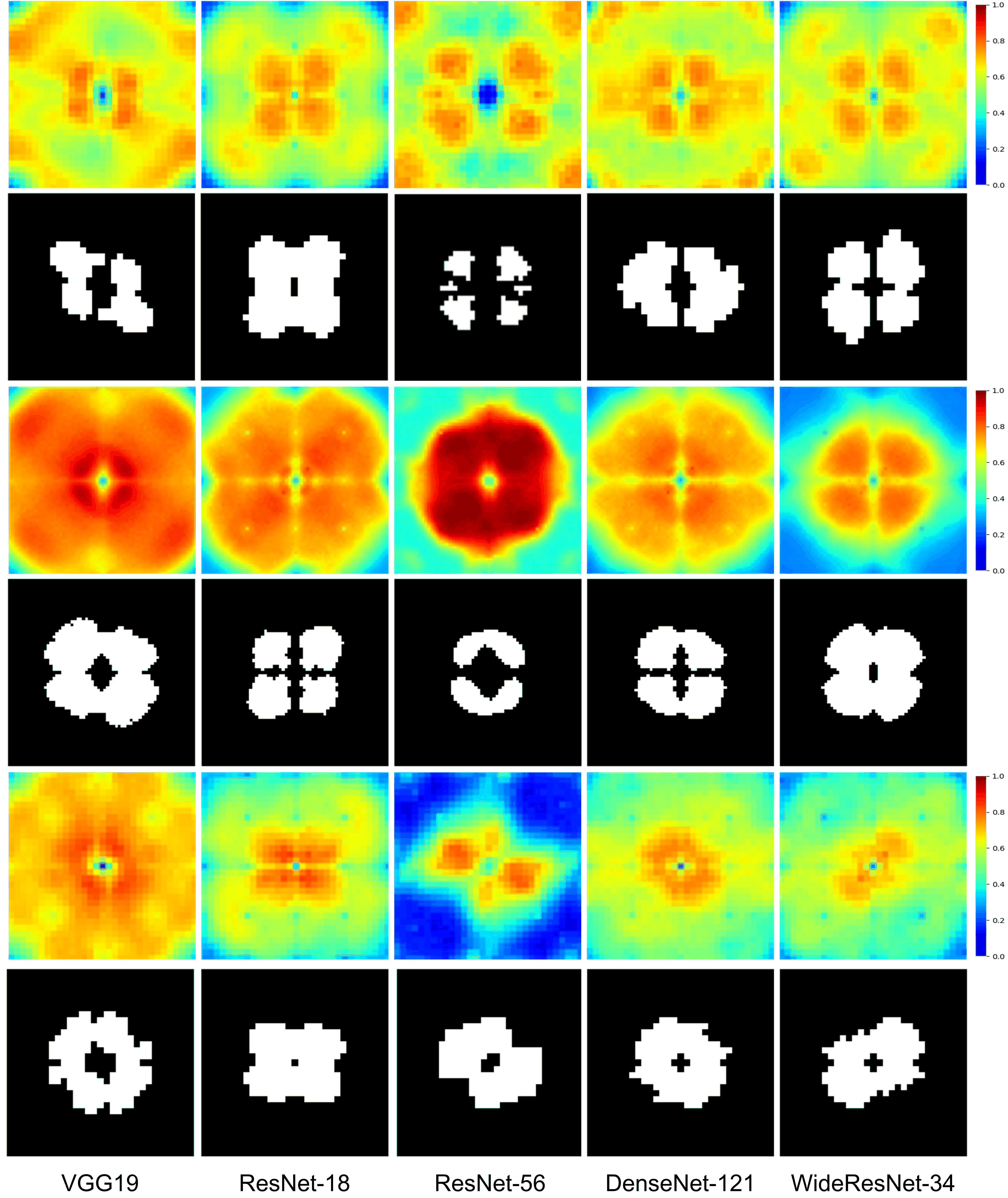}
\caption{Fourier heat maps and mid-low frequency masks for different models on different datasets. The first, third and fifth rows show the Fourier heat maps on CIFAR-10, CIFAR-100 and GTSRB, respectively. The second, fourth and sixth rows the corresponding mid-low frequency masks, respectively. Detailed information about the models and datasets are in the experimental section.}
\end{figure}

\subsubsection{Frequency sensitivity clustering} However, as shown in Fig. 2, different frequencies in the mid-low frequency region result in different test errors. On the one hand, this is surely affected by the image content and training strategy, e.g., the Fourier heat map may reasonably change due to different training parameters even though the frequency principle will still hold. On the other hand, it is required that the watermarking operation should not impair the performance of the DNN on its original task, namely, it is not suggested to perturb those mid-low frequencies that have very low test errors (e.g., the dark-blue region in Fig. 2) because these low-test-error frequencies are very important for the DNN to fit the original task and ``a low test error'' implies that the DNN has adequately fitted the original task. If we modify those frequencies with very low test errors (e.g., the dark-blue region in Fig. 2) during watermark embedding, it forces the DNN to learn the watermarking task, rather than the original task, which will significantly degrade the performance of the DNN on its original task.

Therefore, based on the aforementioned analysis, we conclude that in the mid-low frequency region, it is more desirable to use those frequencies with relatively higher test errors for trigger sample generation because they result in a good trade-off between the watermarking task and the original task. One may empirically choose some mid-low frequencies in specific positions out for trigger sample generation. However, from the point of view of algorithmic design, it is very necessary to find a general way to determine the frequencies suitable for trigger sample generation, which motivates us to propose a frequency sensitivity clustering method in this paper.

Mathematically, let $\textbf{t} \in [0,1]^{h\times w}$ represent the Fourier heat map, for which $t_{i,j}$ is the test error for the frequency position $(i,j)$. It is pointed that  the lowest frequency components have been shifted to the center of the spectrum for better analysis, and $|t_{i,j}| = |t_{i',j'}|$ holds if the two positions $(i,j)$ and $(i',j')$ are symmetrical with respect to the center of the spectrum.

First of all, we determine a \emph{sensitivity map} $\textbf{s} = \{0, 1\}^{h\times w}$
\begin{equation}
s_{i,j} =
\left\{\begin{matrix}
1 & t_{i,j} \geq \rho \\
0 & \text{otherwise},
\end{matrix}\right.
\end{equation}
where $\rho \in [0,1]$ is a threshold that controls the total number of sensitivity frequencies. A frequency is deemed \emph{sensitive} if the corresponding element is ``1'' in the sensitivity map. These sensitivity frequencies, actually, are candidate frequencies for trigger sample generation. For each $s_{i,j}~( = 1)$, we determine a feature vector expressed as
\begin{equation}
\textbf{d}_{i,j} = (d_{i,j,0}, d_{i,j,1}),
\end{equation}
where $d_{i,j,0} = 1 - t_{i,j}$ and
\begin{equation}
d_{i,j,1} = \left [ (i - \frac{h}{2})^2 + (j - \frac{w}{2})^2\right]^{\frac{1}{2}}.
\end{equation}
Here, $d_{i,j,1}$ actually represents the Euclidean distance between $(i,j)$ and the center of the spectrum. Thereafter, we apply K-means clustering to divide all the \emph{sensitive frequencies} into two disjoint subsets (namely, the number of clusters is 2) such that one subset will be used for adding perturbation but the other one will be unchanged. It is noted that the frequency at $(i,j)$ is equivalent to the frequency at the symmetrical position with respect to the center of the spectrum. The clustering bases on the aforementioned feature vector, each component of which should be normalized along the corresponding dimension prior to K-means clustering. For example, after normalization, $d_{i,j,0}$ will become $(d_{i,j,0}- \underset{a,b}{\text{min}}~d_{a,b,0})/(\underset{a,b}{\text{max}}~d_{a,b,0} - \underset{a,b}{\text{min}}~d_{a,b,0})$.

After K-means clustering, two clusters can be obtained. Fig. 3 provides some visual examples, from which we can find that the sensitive frequencies generated by different models on same or different datasets are different from each other, that is, disturbances at different frequencies have different influences on the model. The cluster with the lower average Euclidean distance to the center of the spectrum will be used for trigger sample generation. The other one will be unchanged. The most important advantage of using K-means clustering is that it is \emph{model dependent}, i.e., the generated perturbation considers the influence caused by the model. It is significantly different from many existing methods that directly insert a noticeable signal into a clean sample in the spatial or frequency domain, and makes a step towards \emph{interpretable DNN watermarking}.

According to the aforementioned analysis, we are now ready to describe the steps of trigger sample generation, which are provided in Algorithm 1. Fig. 1 (a) also shows an example for sensitivity map and clustering map. In the clustering map, the white positions constitute the cluster used for trigger sample generation. In the next section, we will conduct experiments and analysis to verify the superiority and applicability.

\emph{Remark:} In Line 6 of Algorithm 1, $\textbf{s}_1'$, $\textbf{s}_2'$, ..., $\textbf{s}_{|S_N|}'$ should use the same perturbation, namely, in Eq. (7), the perturbation term $\lambda_{(i,j),k}\cdot \mathcal{F}(\textbf{e}_{(i,j),k})$ for $\textbf{s}_1'$, $\textbf{s}_2'$, ..., $\textbf{s}_{|S_N|}'$ are equal to each other, which facilitates learning the trigger pattern. Notice that, for a single trigger sample, different sensitive frequencies will use different perturbations controlled by a secret key.

\begin{table}[!t]
\centering
\caption{Relationships between different subsets.}
\scalebox{0.95}{
\begin{tabular}{cc|c|c}
\hline\hline
\multicolumn{2}{c|}{Original dataset} & Subset & Trigger set\\
\hline
\multicolumn{1}{c|}{\multirow{4}{*}{$D$}} & \multirow{2}{*}{$D_1$} & $A_1$ & $B_1$\\
\cline{3-4}
\multicolumn{1}{c|}{} & & $D_1\setminus A_1$ & -\\
\cline{2-4}
\multicolumn{1}{c|}{} & \multirow{2}{*}{$D_2 = D\setminus D_1$} & $A_2$ & $B_2$\\
\cline{3-4}
\multicolumn{1}{c|}{} & & $D_2\setminus A_2$ & -\\
\hline
\multicolumn{2}{c|}{\multirow{2}{*}{$E$}} & $U$ & -\\
\cline{3-4}
& & $V = E\setminus U$ & $V_\text{T}$\\
\hline
\multicolumn{2}{c|}{Other constraints} & \multicolumn{1}{c}{$T_1 = B_1\cup B_2$} & \multicolumn{1}{c}{$T_2 = V_\text{T}$}\\
\hline
\multicolumn{2}{c|}{Dataset for watermark embedding} & \multicolumn{2}{c}{$D\cup T_1 = D_1\cup D_2\cup B_1\cup B_2$}\\
\hline
\multicolumn{2}{c|}{Dataset for ownership verification} & \multicolumn{2}{c}{$T_2 = V_\text{T}$}\\
\hline
\multicolumn{2}{c|}{Dataset for original task evaluation} & \multicolumn{2}{c}{$U\cup V$}\\
\hline\hline
\end{tabular}}
\end{table}

\begin{table}[!t]
\centering
\caption{Classification accuracy on the original task for different models before watermarking.}
\begin{tabular}{c|ccc}
\hline\hline
\multirow{2}{*}{Model} & \multicolumn{3}{c}{Dataset}\\
\cline{2-4}
 & CIFAR-10 & CIFAR-100 & GTSRB\\
\hline
VGG19 & 91.43\% & 67.38\% & 97.31\%\\
ResNet-18 & 93.96\% & 75.76\% & 97.46\%\\
ResNet-56 & 83.85\% & 72.86\% & 98.28\%\\
DenseNet-121 & 94.54\% & 78.63\% & 97.63\%\\
WideResNet-34 & 94.79\% & 79.51\% & 98.56\%\\
\hline\hline
\end{tabular}
\end{table}

\begin{figure*}[!t]
\begin{center}
\includegraphics[width=\linewidth]{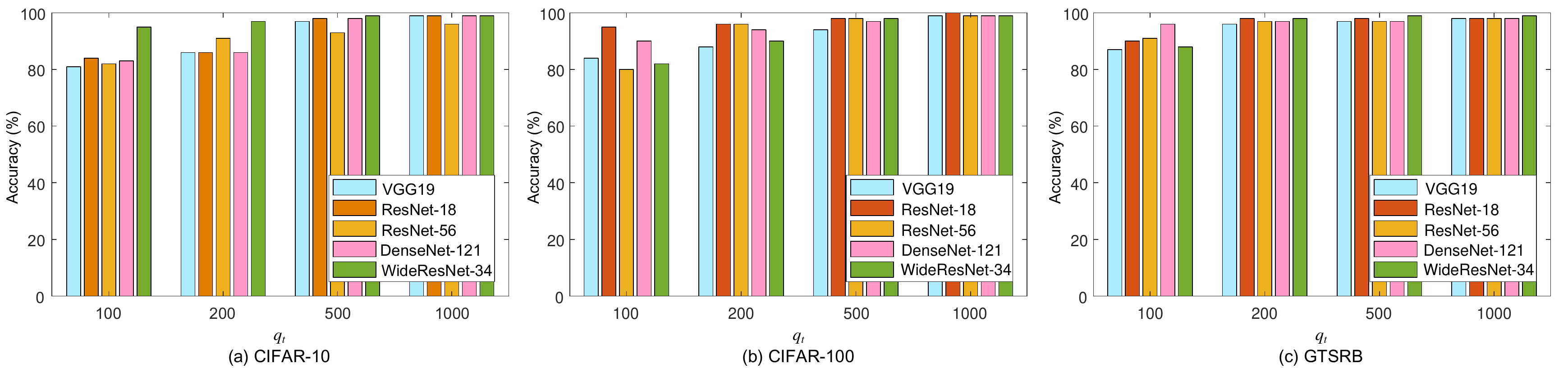}
\caption{Classification accuracy on the trigger set for different models after watermarking with different $q_t$: (a) CIFAR-10, (b) CIFAR-100 and (c) GTSRB.}
\end{center}
\end{figure*}

\section{Experimental Results and Analysis}
\subsection{Setup}
We used three popular datasets CIFAR-10, CIFAR-100 \cite{CIFAR10:paper} and GTSRB \cite{GTSRB:paper} to conduct extensive experiments. CIFAR-10 and CIFAR-100 contain 60,000 images. CIFAR-10 contains 10 classes and CIFAR-100 contains 100 classes. Each of them was randomly divided into three disjoint subsets, i.e., training set (75\%), validation set (5\%) and testing set (20\%). GTSRB contains 43 classes of traffic signs, split into 39,209 training images and 12,630 testing images. The training images were divided into two disjoint subsets, i.e., training set (80\%) and validation set (20\%). The testing images were belonging to the testing set. The popular architectures VGG19 \cite{VGG19}, ResNet-18 \cite{ResNet:paper}, ResNet-56 \cite{ResNet:paper}, DenseNet-121 \cite{DenseNet121}, and WideResNet-34 \cite{wideresnet34} were used to act as the original DNN. The proposed work is not subjected to the above models and datasets.

Without the loss of generalization, we here take VGG19 for example to explain how to generate the trigger samples, how to generate the marked model and how to verify the ownership.

\begin{itemize}
	\item[-] \emph{Trigger Sample Generation:} The training set denoted by $D_1$ and the validation set denoted by $D_2$ are collected to build the dataset $D$ in Algorithm 1. $\mathcal{M}_0$ is set to VGG19. $\mathcal{M}_0$ can be trained with $D$ from scratch to generate the trained but non-marked model (see Line 2 in Algorithm 1). Given any normal sample, we are able to generate the corresponding trigger sample by Algorithm 1.
	\item[-] \emph{Watermark Embedding:} We randomly generate two sets $A_1\subset D_1$ and $A_2\subset D_2$. We apply Algorithm 1 to $A_1$ and $A_2$ to generate the corresponding trigger samples, whose labels are set to $c$. In this way, we obtain two trigger sets $B_1$ (corresponding to $A_1$) and $B_2$ (corresponding to $A_2$). Two sets $D_1\cup B_1$ (training) and $D_2\cup B_2$ (validation) are used to train $\mathcal{M}_0$ from scratch to generate the trained and marked model $\mathcal{M}_1$. It can be inferred that $T_1 = B_1\cup B_2$ according to the definition of $T_1$ in Subsection II-B.
	\item[-] \emph{Ownership Verification:} We verify the ownership of the target model by Eq. (4). The key step is how to generate the trigger set $T_2$. To deal with this issue, we divide the testing set $E$ into two disjoint subsets $U$ and $V$. $V$ is used to generate a set of trigger samples $V_\text{T}$. We set $T_2 = V_\text{T}$. Table I shows the relationships between different sets.
\end{itemize}

We set $|A_1|$ = $|B_1|$ = $|A_2|$ = $|B_2|$ = $|V|$ = $|V_T|$ = $q_t$. We empirically set $q_t = 500$ by default in our experiments unless otherwise specified. We applied the stochastic gradient descent (SGD) optimizer to train each network. The learning rate was 0.01, with a momentum of 0.9. The batch size was 512. The perturbation coefficients were randomly sampled from $[-1, 1]$. The threshold $\rho$ was set to 0.65 by default. It is always free for us to adjust these parameters to achieve better performance.

\begin{table}[!t]
\centering
\caption{Classification accuracy on the original task for different models after watermarking.}
\scalebox{0.95}{
\begin{tabular}{c|c|cccc}
\hline\hline
\multirow{2}{*}{Dataset} & \multirow{2}{*}{Model} & \multicolumn{4}{c}{$q_t$}\\
\cline{3-6}
 & & 100 & 200 & 500 & 1000\\
\hline
& VGG19 & 91.43\% & 91.36\% & 91.32\% & 90.87\%\\
& ResNet-18 & 93.95\% & 93.91\% & 93.88\% & 93.74\%\\
CIFAR-10 & ResNet-56 & 83.86\% & 83.79\% & 83.41\% & 83.36\%\\
& DenseNet-121 & 94.54\% & 94.52\% & 94.48\% & 93.78\%\\
& WideResNet-34 & 94.82\% & 94.86\% & 94.53\% & 94.36\%\\
\hline
& VGG19 & 67.37\% & 67.37\% & 67.34\% & 67.28\%\\
& ResNet-18 & 75.78\% & 75.76\% & 75.72\% & 75.67\%\\
CIFAR-100 & ResNet-56 & 72.86\% & 72.84\% & 72.81\% & 72.76\%\\
& DenseNet-121 & 78.62\% & 78.54\% & 78.43\% & 78.29\%\\
& WideResNet-34 & 79.50\% & 79.49\% & 78.71\% & 78.46\%\\
\hline
& VGG19 & 97.34\% & 97.30\% & 97.28\% & 97.24\%\\
& ResNet-18 & 97.45\% & 97.46\% & 96.78\% & 96.63\%\\
GTSRB & ResNet-56 & 98.30\% & 98.27\% & 98.23\% & 97.56\%\\
& DenseNet-121 & 97.61\% & 97.62\% & 97.57\% & 98.38\%\\
& WideResNet-34 & 98.57\% & 98.59\% & 98.51\% & 98.85\%\\
\hline\hline
\end{tabular}}
\end{table}

\begin{figure}[!t]
\begin{center}
\includegraphics[width=\linewidth]{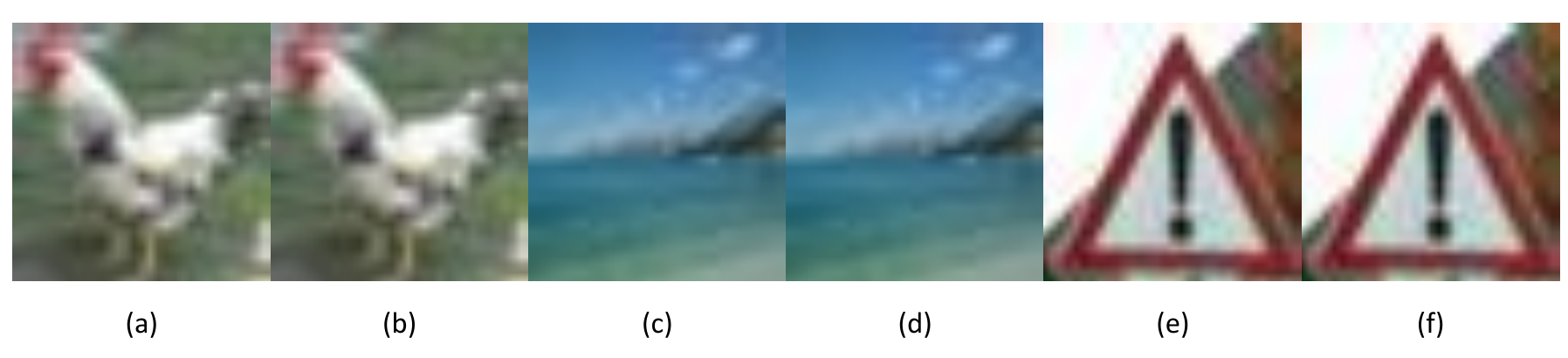}
\caption{Examples of the perturbed sample: (a, c, e) original examples randomly selected from CIFAR-10, CIFAR-100 and GTSRB, respectively, (b, d, f) the corresponding perturbed samples. The representative ResNet-18 was used.}
\end{center}
\end{figure}

\begin{figure}[!t]
\begin{center}
\includegraphics[width=3in]{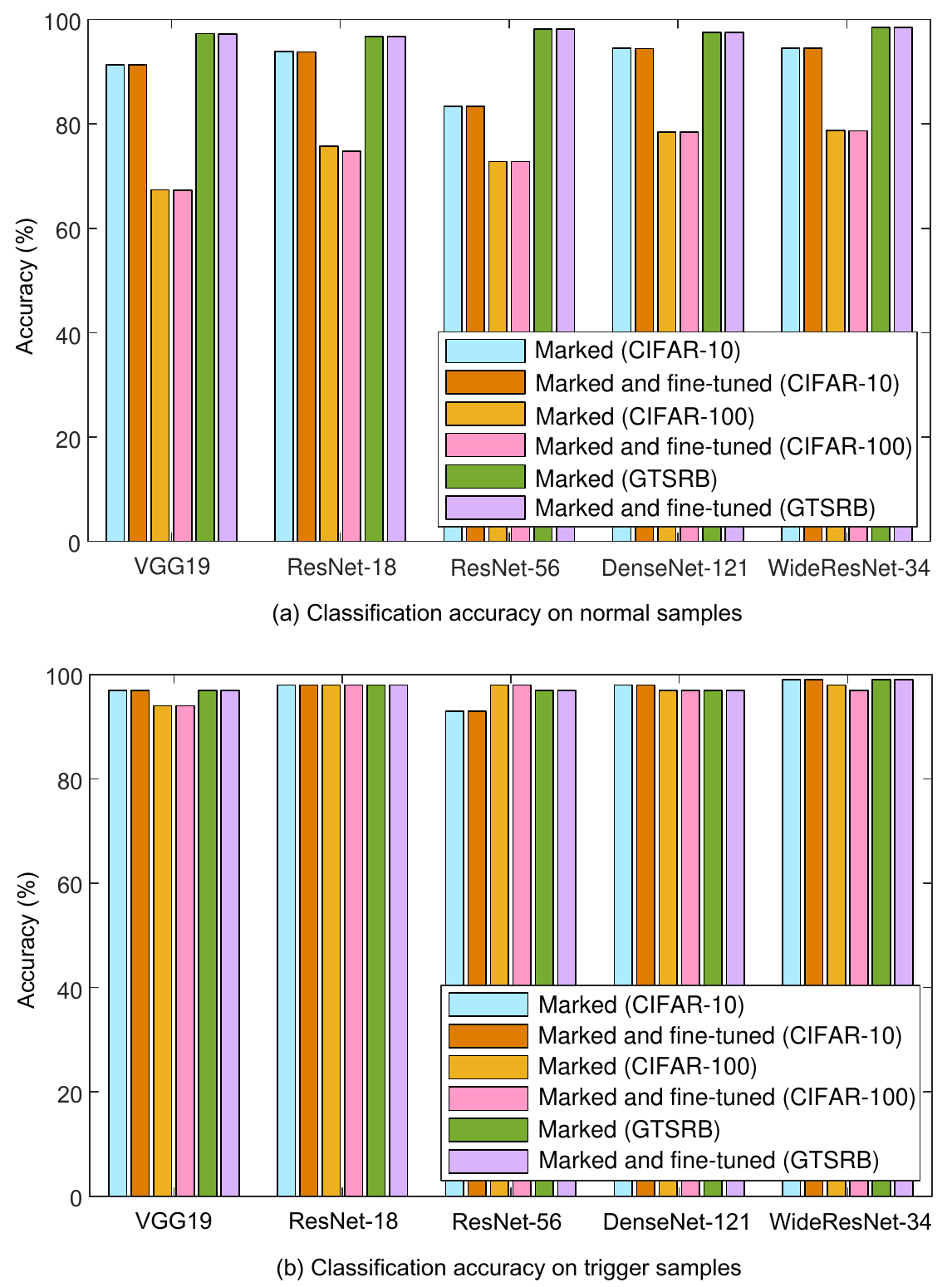}
\caption{Performance evaluation after model fine-tuning: (a) classification accuracy on normal samples, (b) classification accuracy on trigger samples.}
\end{center}
\end{figure}

\begin{figure*}[!t]
\begin{center}
\includegraphics[width=\linewidth]{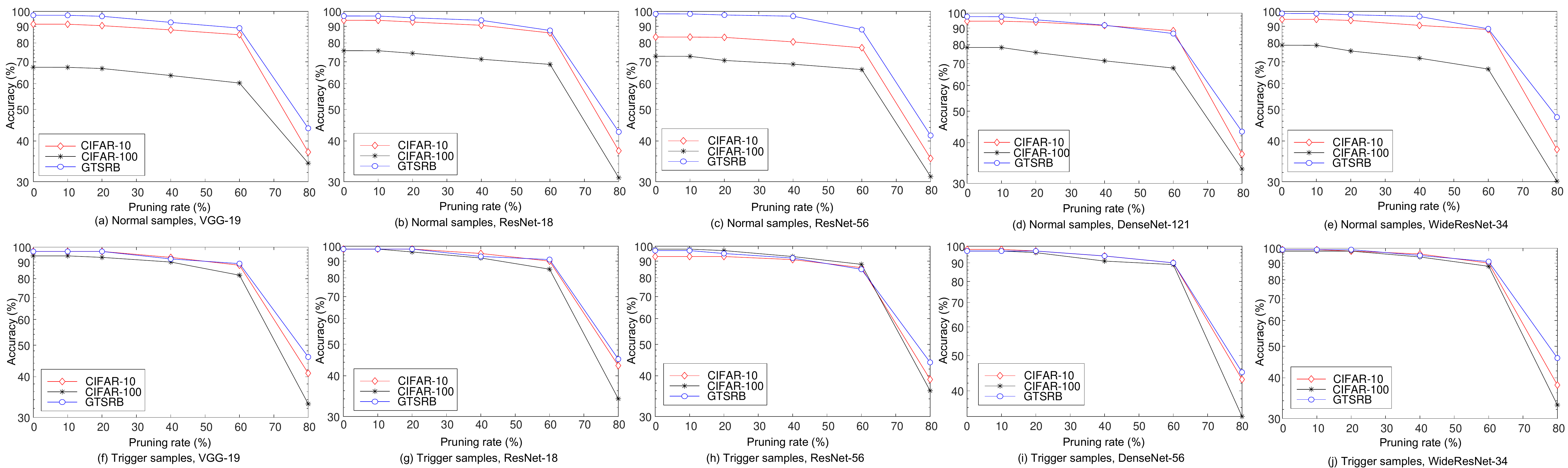}
\caption{Performance evaluation after model pruning: (a-e) classification accuracy on normal samples, (f-j) classification accuracy on trigger samples.}
\end{center}
\end{figure*}

\subsection{Fidelity}
Task fidelity means that the performance of the DNN model on its original task after watermarking should be kept well. It requires us to determine the accuracy of image classification for the original model (which is deemed \emph{non-marked}) and the \emph{marked} model. Table II shows the classification accuracy of the non-marked model. Table III shows the classification accuracy of the marked model. It can be inferred from Table II that different models have different performance on different image classification tasks, which is reasonable due to the different learning capabilities of deep models. It can be also inferred from Table III that the accuracy declines as $q_t$ increases from the viewpoint of overall trend, which is a normal phenomenon because the model needs to make a sacrifice on the original task for model watermarking. However, by comparing Table II and Table III, we can find that the performance degradation on the original task is very low after watermarking with different $q_t$, which indicates that the proposed method does not impair the utilization of the model and thereby has good potential in application scenarios. It is admitted that we did not optimize the hyper-parameters and the training strategy for all models, meaning that the baseline performance in Table II may be not optimal, which does not affect the evaluation of our work.

\begin{table}[!t]
\centering
\caption{The mean PSNR and the mean SSIM between the original samples and the corresponding perturbed samples. The representative ResNet-18 was used as the host network.}
\begin{tabular}{c|ccc}
\hline\hline
\multirow{2}{*}{Indicator} & \multicolumn{3}{c}{Dataset}\\
\cline{2-4}
 & CIFAR-10 & CIFAR-100 & GTSRB\\
\hline
Mean PSNR & 36.14 dB & 37.46 dB & 38.53 dB\\
Mean SSIM & 0.9862   & 0.9927   & 0.9935\\
\hline\hline
\end{tabular}
\end{table}

On the other hand, watermark fidelity means that the hidden watermark should be reconstructed with a high accuracy. Since the proposed method corresponds to zero-bit watermarking, it is required that the classification accuracy on the trigger set should be as high as possible. We calculate the classification accuracy on the trigger set for different models with different $q_t$. The results are shown in Fig. 4, from which we can find that different models have different accuracy values, which means that different models have different capabilities of carrying a watermark. It is observed that all the accuracy values are no less than 80\%, meaning that the proposed method enables the watermark to be reliably extracted. It can be inferred that the accuracy will increase when $q_t$ increases in most cases. The reason is that a larger $q_t$ means that more trigger samples are used for model training, allowing the watermarking task to be accomplished very well by the model. For example, in Fig. 4, the accuracy is approaching 100\% when $q_t = 1000$, which has demonstrated the superiority of the proposed method.

In addition, sample fidelity means that the perturbed sample should be visually close to the original sample. Fig. 5 shows some examples of the perturbed sample. It can be seen that the proposed perturbation technique does not introduce noticeable artifacts. To quantize the visual quality, we determine the mean peak signal-to-noise ratio (PSNR, dB) and structural similarity (SSIM) \cite{Wang:TIP:2004} between the original samples in the testing set and the corresponding perturbed samples. The results are shown in Table IV, from which we can find that the mean PSNRs are higher than 36 dB and the mean SSIMs are higher than 0.99. It indicates that the perturbed samples have very good visual quality, which can guarantee perturbation imperceptibility.

\subsection{Robustness}
Generally, robustness evaluates the ability to reconstruct the embedded watermark from the marked DNN model when the watermarking system was attacked by the adversary. Unlike many existing methods only considering attacks to the marked model, we further take into account the possible attacks to the trigger samples. In the following, we first analyze the performance of the model on the original task and the watermarking task when the marked model was attacked. Then, we analyze the performance when the trigger samples were attacked.

Two most popular attacks are applied to the marked model. One is model fine-tuning, and the other is model pruning. To mimic the fine-tuning attack, we randomly choose 50\% normal samples of the validation set to fine-tune the marked model. To mimic the model pruning attack, we apply the $\ell_1$-norm pruning strategy to the marked model, where a pruning rate is used to denote the percentage of pruned parameters. We evaluate the performance of the attacked model on the original task and the watermarking task. The results are provided in Fig. 6 and Fig. 7. It is inferred from Fig. 6 that model fine-tuning does not impair the original task and the watermarking task. For model pruning, though the performance will gradually decline as the pruning rate increases, the performance can be well preserved even 50\% of the parameters are pruned, which indicates that the proposed method has good ability to resist model pruning.

We further evaluate the ability of the trigger samples against image processing attacks. As the computing resource is limited, only the representative VGG19 and ResNet-18 are used as the host network unless otherwise specified. We consider three common attacks, i.e., JPEG compression, flipping and filtering. A common approach to enhance the robustness against these attacks is using data augmentation (UDA), i.e., mixing the attacked images into the training set. Both UDA and not using data augmentation (NUDA) are considered here for fair evaluation. For UDA, to resist a specific attack with preseted parameter(s), we enhance the robustness and generalization of the model by putting the normal training samples processed by the same attack into the training set. Thus, the model not only learns the original task and the watermarking task well, but also resists the corresponding attack applied to the input.

\begin{table*}[!t]
\centering
\caption{The performance against JPEG compression under different conditions for VGG19. $\text{Acc}_\text{o}$ represents the classification accuracy for the original task and $\text{Acc}_\text{w}$ represents the classification accuracy for the watermarking task.}
\begin{tabular}{c|c|cc|cc|cc|cc|cc}
\hline\hline
\multirow{2}{*}{Dataset}  & Quality factor & \multicolumn{2}{c|}{100}           & \multicolumn{2}{c|}{80}            & \multicolumn{2}{c|}{60}            & \multicolumn{2}{c|}{40}            & \multicolumn{2}{c}{20}              \\ \cline{2-12}
                          & Indicator      & \multicolumn{1}{c|}{$\text{Acc}_\text{o}$} &$\text{Acc}_\text{w}$&\multicolumn{1}{c|}{$\text{Acc}_\text{o}$} & $\text{Acc}_\text{w}$ & \multicolumn{1}{c|}{$\text{Acc}_\text{o}$} & $\text{Acc}_\text{w}$ & \multicolumn{1}{c|}{$\text{Acc}_\text{o}$} & $\text{Acc}_\text{w}$ & \multicolumn{1}{c|}{$\text{Acc}_\text{o}$}  & $\text{Acc}_\text{w}$ \\ \hline
\multirow{2}{*}{CIFAR-10}  & NUDA           & \multicolumn{1}{c|}{91.32\%} & 97\% & \multicolumn{1}{c|}{87.95\%} & 87\% & \multicolumn{1}{c|}{83.76\%} & 35\%& \multicolumn{1}{c|}{75.64\%} & 3\% & \multicolumn{1}{c|}{61.28\%} & 0\% \\ \cline{2-12}
                          & UDA            & \multicolumn{1}{c|}{91.23\%} & 97\% & \multicolumn{1}{c|}{86.68\%} & 96\% & \multicolumn{1}{c|}{82.51\%} & 71\% & \multicolumn{1}{c|}{73.79\%} & 7\% & \multicolumn{1}{c|}{60.83\%}  & 0\% \\ \hline
\multirow{2}{*}{CIFAR-100} & NUDA           & \multicolumn{1}{c|}{67.34\%} & 95\% & \multicolumn{1}{c|}{65.78\%} & 82\% & \multicolumn{1}{c|}{52.83\%} & 22\% & \multicolumn{1}{c|}{46.95\%} & 2\% & \multicolumn{1}{c|}{42.76\%} & 0\% \\ \cline{2-12}
                          & UDA            & \multicolumn{1}{c|}{67.16\%} & 95\% & \multicolumn{1}{c|}{65.23\%} & 93\% & \multicolumn{1}{c|}{50.42\%} & 68\% & \multicolumn{1}{c|}{43.48\%} & 5\% & \multicolumn{1}{c|}{41.32\%}  & 0\% \\ \hline
\multirow{2}{*}{GTRSB}    & NUDA           & \multicolumn{1}{c|}{97.28\%} & 98\% & \multicolumn{1}{c|}{96.46\%} & 85\% & \multicolumn{1}{c|}{88.89\%} & 28\% & \multicolumn{1}{c|}{72.37\%} & 3\% & \multicolumn{1}{c|}{67.06\%} & 0\% \\ \cline{2-12}
                          & UDA            & \multicolumn{1}{c|}{97.05\%} & 98\% & \multicolumn{1}{c|}{96.37\%} & 98\% & \multicolumn{1}{c|}{88.78\%} & 74\% & \multicolumn{1}{c|}{70.96\%} & 8\% & \multicolumn{1}{c|}{66.85\%}  & 0\% \\ \hline\hline
\end{tabular}
\end{table*}

\begin{table*}[!t]
\centering
\caption{The performance against JPEG compression under different conditions for ResNet-18.}
\begin{tabular}{c|c|cc|cc|cc|cc|cc}
\hline\hline
\multirow{2}{*}{Dataset}   & Quality factor & \multicolumn{2}{c|}{100}           & \multicolumn{2}{c|}{80}            & \multicolumn{2}{c|}{60}            & \multicolumn{2}{c|}{40}            & \multicolumn{2}{c}{20}              \\ \cline{2-12}
                           & Indicator      & \multicolumn{1}{c|}{$\text{Acc}_\text{o}$} &$\text{Acc}_\text{w}$&\multicolumn{1}{c|}{$\text{Acc}_\text{o}$} &$\text{Acc}_\text{w}$&\multicolumn{1}{c|}{$\text{Acc}_\text{o}$} &$\text{Acc}_\text{w}$&\multicolumn{1}{c|}{$\text{Acc}_\text{o}$} &$\text{Acc}_\text{w}$&\multicolumn{1}{c|}{$\text{Acc}_\text{o}$}  & $\text{Acc}_\text{w}$ \\ \hline
\multirow{2}{*}{CIFAR-10}  & NUDA           & \multicolumn{1}{c|}{93.88\%} & 99\% & \multicolumn{1}{c|}{89.96\%} & 88\% & \multicolumn{1}{c|}{84.65\%} & 29\% & \multicolumn{1}{c|}{78.91\%} & 3\% & \multicolumn{1}{c|}{68.57\%}  & 0\% \\ \cline{2-12}
                           & UDA            & \multicolumn{1}{c|}{93.69\%} & 99\% & \multicolumn{1}{c|}{89.75\%} & 99\% & \multicolumn{1}{c|}{84.33\%} & 76\% & \multicolumn{1}{c|}{77.42\%} & 6\% & \multicolumn{1}{c|}{66.94\%} & 0\% \\ \hline
\multirow{2}{*}{CIFAR-100} & NUDA           & \multicolumn{1}{c|}{75.72\%} & 99\% & \multicolumn{1}{c|}{73.87\%} & 87\% & \multicolumn{1}{c|}{67.49\%} & 23\% & \multicolumn{1}{c|}{58.70\%} & 5\% & \multicolumn{1}{c|}{53.42\%}  & 0\% \\ \cline{2-12}
                           & UDA            & \multicolumn{1}{c|}{75.51\%} & 99\% & \multicolumn{1}{c|}{72.28\%} & 96\% & \multicolumn{1}{c|}{63.91\%} & 63\% & \multicolumn{1}{c|}{55.54\%} & 7\% & \multicolumn{1}{c|}{52.85\%} & 0\% \\ \hline
\multirow{2}{*}{GTRSB}     & NUDA           & \multicolumn{1}{c|}{96.78\%} & 99\% & \multicolumn{1}{c|}{94.88\%} & 89\% & \multicolumn{1}{c|}{89.27\%} & 37\% & \multicolumn{1}{c|}{75.65\%} & 6\% & \multicolumn{1}{c|}{65.88\%}  & 0\% \\ \cline{2-12}
                           & UDA            & \multicolumn{1}{c|}{96.63\%} & 99\% & \multicolumn{1}{c|}{94.64\%} & 99\% & \multicolumn{1}{c|}{87.49\%} & 78\% & \multicolumn{1}{c|}{74.87\%} & 9\% & \multicolumn{1}{c|}{64.79\%} & 0\% \\ \hline\hline
\end{tabular}
\end{table*}

Table V and Table VI show the performance against JPEG compression for VGG19 and ResNet-18, respectively. It can be inferred that for both UDA and NUDA, when the quality factor (QF) gradually decreases, both $\text{Acc}_\text{o}$ and $\text{Acc}_\text{w}$ tend to decrease, which is due to the reason that a smaller QF results in more loss of feature information. However, we can find that the performance difference between UDA and NUDA is small, especially for larger QFs. It indicates that the proposed method has good ability to make the trigger samples capable of resisting JPEG compression. It is noted that both the normal sample and the trigger samples are attacked during the testing phase. Table VII provides the results against horizontal flipping under different conditions, from which we can infer that flipping does not impair watermarking because the trigger patterns are symmetrical in
the frequency domain.

\begin{table}[!t]
\caption{The performance against horizontal flipping.}
\centering
\begin{tabular}{c|c|cc|cc}
\hline\hline
\multirow{2}{*}{Dataset} & Model & \multicolumn{2}{c|}{VGG19} & \multicolumn{2}{c}{ResNet-18} \\ \cline{2-6}

 & Indicator & \multicolumn{1}{c|}{Acc$_\text{o}$} & Acc$_\text{w}$ & \multicolumn{1}{c|}{Acc$_\text{o}$} & Acc$_\text{w}$ \\ \hline

\multirow{2}{*}{CIFAR-10}  & NUDA & \multicolumn{1}{c|}{91.32\%} & 97\% & \multicolumn{1}{c|}{93.88\%} & 99\% \\
                          & UDA & \multicolumn{1}{c|}{91.15\%} & 97\% & \multicolumn{1}{c|}{93.75\%} & 99\% \\ \hline
\multirow{2}{*}{CIFAR-100} & NUDA & \multicolumn{1}{c|}{67.34\%} & 95\% & \multicolumn{1}{c|}{75.72\%} & 99\% \\
                          & UDA & \multicolumn{1}{c|}{66.29\%} & 95\% & \multicolumn{1}{c|}{75.38\%} & 99\% \\ \hline
\multirow{2}{*}{GTRSB}    & NUDA & \multicolumn{1}{c|}{97.28\%} & 98\% & \multicolumn{1}{c|}{96.78\%} & 99\% \\
                          & UDA & \multicolumn{1}{c|}{97.17\%} & 98\% & \multicolumn{1}{c|}{96.43\%} & 99\% \\ \hline\hline
\end{tabular}
\end{table}

We further consider low-pass filtering in the DFT domain. Low-pass filtering with bandwidth B is defined as the operation of setting all frequency components outside the center square of width B in the Fourier spectrum at the center lowest frequency to zero and then applying the inverse DFT to the spectrum. Table VIII and Table IX report the performance against low-pass filtering under different conditions for VGG19 and ResNet-18. It can be inferred that the classification accuracy on the trigger set achieves more than 80\% when the bandwidth is larger than 12. However, when the bandwidth is smaller than 12, it causes a significant drop of the classification accuracy for both the original mission and the watermarking task. This is because our method perturbs mid-low frequency components. When the trigger samples pass through a large filter, although some high-frequency information is lost, the model can still learn the trigger. The image content can be learned by the model as well. Therefore, both the original task and the watermarking task are performed well. However, by applying a smaller filter, much information about the trigger and the image content will be lost. As a result, the performance on both tasks will be surely declined.

\begin{table*}[!t]
\centering
\caption{The performance against low-pass filtering under different conditions for VGG19.}
\begin{tabular}{c|c|cc|cc|cc|cc|cc}
\hline\hline
\multirow{2}{*}{Dataset}  & Bandwidth & \multicolumn{2}{c|}{16} & \multicolumn{2}{c|}{14} & \multicolumn{2}{c|}{12} & \multicolumn{2}{c|}{8} & \multicolumn{2}{c}{4} \\ \cline{2-12}
                          & Indicator & \multicolumn{1}{c|}{$\text{Acc}_\text{o}$} & $\text{Acc}_\text{w}$ & \multicolumn{1}{c|}{$\text{Acc}_\text{o}$} & $\text{Acc}_\text{w}$ & \multicolumn{1}{c|}{$\text{Acc}_\text{o}$} & $\text{Acc}_\text{w}$ & \multicolumn{1}{c|}{$\text{Acc}_\text{o}$} & $\text{Acc}_\text{w}$ & \multicolumn{1}{c|}{$\text{Acc}_\text{o}$}  & $\text{Acc}_\text{w}$ \\ \hline
\multirow{2}{*}{CIFAR10}  & NUDA      & \multicolumn{1}{c|}{91.32\%} & 97\% & \multicolumn{1}{c|}{82.85\%} & 87\% & \multicolumn{1}{c|}{73.83\%} & 82\% & \multicolumn{1}{c|}{36.19\%} & 23\% & \multicolumn{1}{c|}{23.83\%}  & 0\%  \\ \cline{2-12}
                          & UDA       & \multicolumn{1}{c|}{91.18\%} & 97\% & \multicolumn{1}{c|}{81.36\%} & 92\% & \multicolumn{1}{c|}{72.67\%} & 83\% & \multicolumn{1}{c|}{34.37\%} & 24\% & \multicolumn{1}{c|}{21.26\%}  & 0\%  \\ \hline
\multirow{2}{*}{CIFAR100} & NUDA      & \multicolumn{1}{c|}{67.34\%} & 95\% & \multicolumn{1}{c|}{62.28\%} & 81\% & \multicolumn{1}{c|}{49.45\%} & 72\% & \multicolumn{1}{c|}{27.49\%} & 16\% & \multicolumn{1}{c|}{13.38\%}  & 0\%  \\ \cline{2-12}
                          & UDA       & \multicolumn{1}{c|}{67.21\%} & 95\% & \multicolumn{1}{c|}{61.73\%} & 87\% & \multicolumn{1}{c|}{47.64\%} & 78\% & \multicolumn{1}{c|}{25.53\%} & 19\% & \multicolumn{1}{c|}{11.59\%}  & 0\%  \\ \hline
\multirow{2}{*}{GTRSB}    & NUDA      & \multicolumn{1}{c|}{97.28\%} & 98\% & \multicolumn{1}{c|}{87.79\%} & 81\% & \multicolumn{1}{c|}{75.69\%} & 81\% & \multicolumn{1}{c|}{46.38\%} & 25\% & \multicolumn{1}{c|}{19.06\%} & 0\%  \\ \cline{2-12}
                          & UDA       & \multicolumn{1}{c|}{97.13\%} & 98\% & \multicolumn{1}{c|}{86.49\%} & 95\% & \multicolumn{1}{c|}{74.91\%} & 85\% & \multicolumn{1}{c|}{45.72\%} & 28\% & \multicolumn{1}{c|}{18.76\%}  & 0\%  \\ \hline\hline
\end{tabular}
\end{table*}

\begin{table*}[!t]
\centering
\caption{The performance against low-pass filtering under different conditions for ResNet-18.}
\begin{tabular}{c|c|cc|cc|cc|cc|cc}
\hline\hline
\multirow{2}{*}{Dataset} & Bandwidth & \multicolumn{2}{c|}{16} & \multicolumn{2}{c|}{14} & \multicolumn{2}{c|}{12}  & \multicolumn{2}{c|}{8} & \multicolumn{2}{c}{4}  \\ \cline{2-12}
                          & Indicator & \multicolumn{1}{c|}{$\text{Acc}_\text{o}$} &$\text{Acc}_\text{w}$&\multicolumn{1}{c|}{$\text{Acc}_\text{o}$} &$\text{Acc}_\text{w}$&\multicolumn{1}{c|}{$\text{Acc}_\text{o}$} &$\text{Acc}_\text{w}$&\multicolumn{1}{c|}{$\text{Acc}_\text{o}$} &$\text{Acc}_\text{w}$&\multicolumn{1}{c|}{$\text{Acc}_\text{o}$}  & $\text{Acc}_\text{w}$ \\ \hline
\multirow{2}{*}{CIFAR10}  & NUDA      & \multicolumn{1}{c|}{93.88\%} & 99\% & \multicolumn{1}{c|}{84.87\%} & 86\% & \multicolumn{1}{c|}{76.86\%} & 80\% & \multicolumn{1}{c|}{43.51\%} & 17\% & \multicolumn{1}{c|}{26.39\%} & 0\%  \\ \cline{2-12}
                          & UDA       & \multicolumn{1}{c|}{93.63\%} & 99\% & \multicolumn{1}{c|}{84.65\%} & 90\% & \multicolumn{1}{c|}{76.57\%} & 84\% & \multicolumn{1}{c|}{41.32\%} & 26\% & \multicolumn{1}{c|}{25.44\%}  & 0\%  \\ \hline
\multirow{2}{*}{CIFAR100} & NUDA      & \multicolumn{1}{c|}{75.72\%} & 99\% & \multicolumn{1}{c|}{70.80\%} & 85\% & \multicolumn{1}{c|}{56.84\%} & 73\% & \multicolumn{1}{c|}{38.79\%} & 15\% & \multicolumn{1}{c|}{19.57\%} & 0\%  \\ \cline{2-12}
                          & UDA       & \multicolumn{1}{c|}{75.45\%} & 99\% & \multicolumn{1}{c|}{69.95\%} & 94\% & \multicolumn{1}{c|}{52.91\%} & 81\% & \multicolumn{1}{c|}{34.37\%} & 28\% & \multicolumn{1}{c|}{17.16\%}  & 0\%  \\ \hline
\multirow{2}{*}{GTRSB}    & NUDA      & \multicolumn{1}{c|}{96.78\%} & 99\% & \multicolumn{1}{c|}{86.83\%} & 89\% & \multicolumn{1}{c|}{78.58\%} & 77\% & \multicolumn{1}{c|}{51.43\%} & 24\% & \multicolumn{1}{c|}{27.92\%} & 0\%  \\ \cline{2-12}
                          & UDA       & \multicolumn{1}{c|}{96.57\%} & 99\% & \multicolumn{1}{c|}{86.38\%} & 96\% & \multicolumn{1}{c|}{77.43\%} & 82\% & \multicolumn{1}{c|}{50.92\%} & 35\% & \multicolumn{1}{c|}{26.35\%}  & 0\%  \\ \hline\hline
\end{tabular}
\end{table*}

\begin{figure}[!t]
\begin{center}
\includegraphics[width=\linewidth]{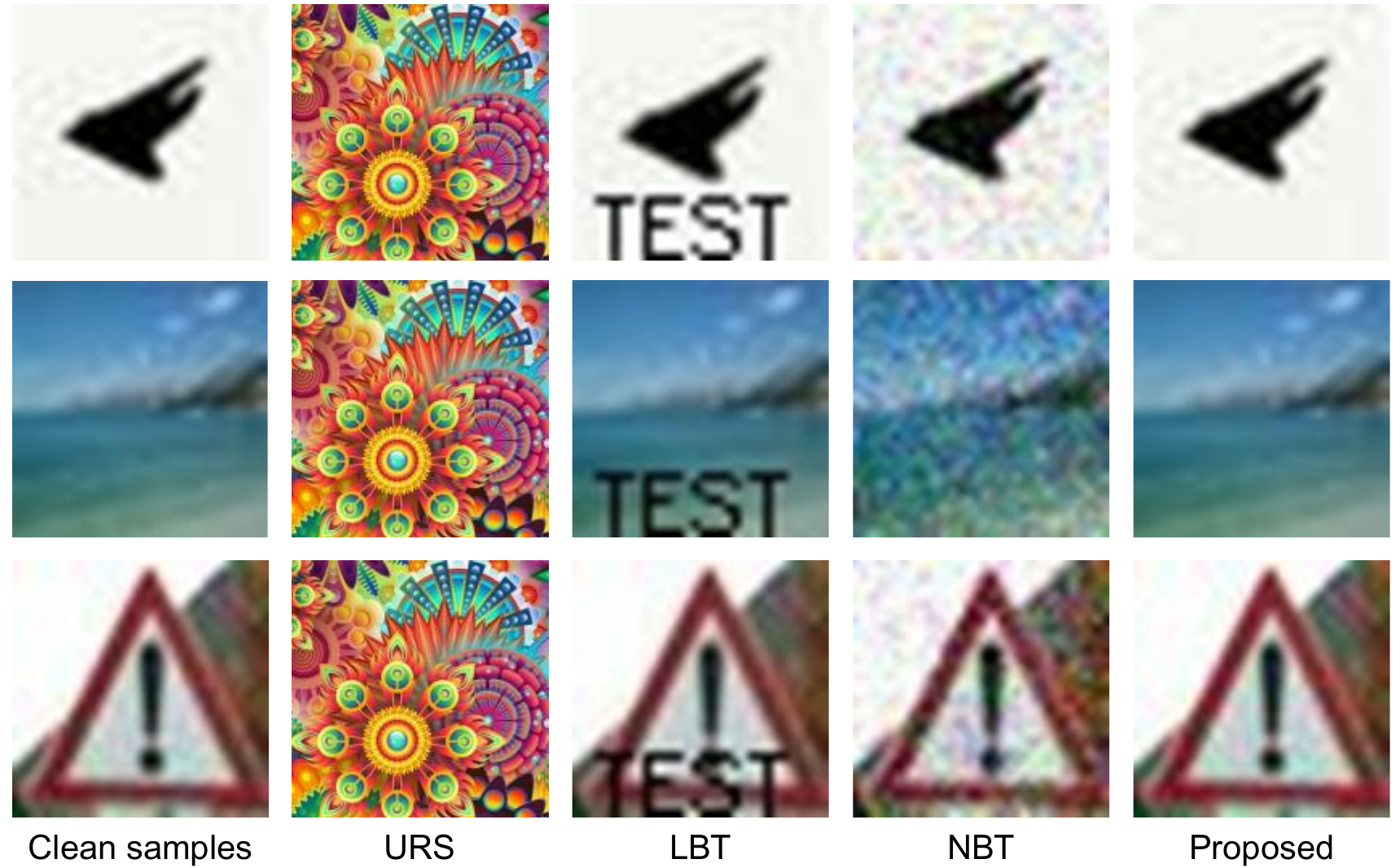}
\caption{Trigger samples by different methods for different datasets (from top to bottom: CIFAR-10, CIFAR-100, GTRSB). Here, NBT uses Gaussian noise with zero mean and a variance of 0.01.}
\end{center}
\end{figure}

\begin{table}[!t]
\centering
\caption{Comparison between different trigger generation strategies for VGG-19. PSNR and SSIM use mean values.}
\begin{tabular}{c|c|c|c|c|c}
\hline\hline
Dataset & Strategy & $\text{Acc}_\text{o}$ & $\text{Acc}_\text{w}$ & PSNR & SSIM\\
\hline
& Non-marked & 91.43\% & - & - & - \\
& URS & 90.98\% & 94\%& - & - \\
CIFAR-10 &  LBT & 90.81\% & 93\% & 12.87& 0.7875 \\
& NBT & 91.08\% & 95\% &24.89 & 0.9073 \\
& Proposed & 91.32\% & 97\% & 38.85 & 0.9884 \\
\hline
& Non-marked & 67.38\% & - & - & - \\
& URS & 66.05\% & 93\% & - & - \\
CIFAR-100 & LBT &65.43\% & 91\%& 12.87 & 0.7875 \\
& NBT &66.17\% & 92\% & 24.89 & 0.9073 \\
& Proposed & 67.34\% & 95\% & 36.62  & 0.9851 \\
\hline
& Non-marked & 97.31\% & - & - & - \\
& URS & 97.10\% & 96\% &- &- \\
GTRSB & LBT &96.87\% &96\% & 12.87 & 0.7875 \\
& NBT & 97.21\% & 97\% & 24.89 & 0.9073 \\
& Proposed & 97.28\% & 98\% & 36.33  & 0.9838 \\
\hline\hline
\end{tabular}
\end{table}

\begin{table}[!t]
\centering
\caption{Comparison between different trigger generation strategies for ResNet-18. PSNR and SSIM use mean values.}
\begin{tabular}{c|c|c|c|c|c}
\hline\hline
Dataset & Strategy & $\text{Acc}_\text{o}$ & $\text{Acc}_\text{w}$ & PSNR & SSIM\\
\hline
& Non-marked & 93.96\% & - & - & - \\
& URS &93.60 &98\% & - & - \\
CIFAR-10 & LBT &92.97 &96\% & 12.87 & 0.7875 \\
& NBT &93.72\% & 96\% & 24.89 & 0.9073 \\
& Proposed & 93.88\% & 99\% & 36.14 & 0.9862 \\
\hline
& Non-marked & 75.76\% & - & - & - \\
& URS &74.29\% & 96\% & - & - \\
CIFAR-100 & LBT & 73.60\% & 94\% & 12.87 & 0.7875 \\
& NBT &74.38\% & 97\% & 24.89 & 0.9073 \\
& Proposed & 75.72\% & 99\% & 37.46 & 0.9927 \\
\hline
& Non-marked & 97.48\% & - & - & - \\
& URS & 96.52\% & 98\% & - & - \\
GTRSB & LBT & 95.85\% & 97\% & 12.87 & 0.7875 \\
& NBT & 96.43\% & 98\% & 24.89 & 0.9073 \\
& Proposed & 96.78\% & 99\% & 38.53 & 0.9935 \\
\hline\hline
\end{tabular}
\end{table}

\begin{figure}[!t]
\begin{center}
\includegraphics[width=\linewidth]{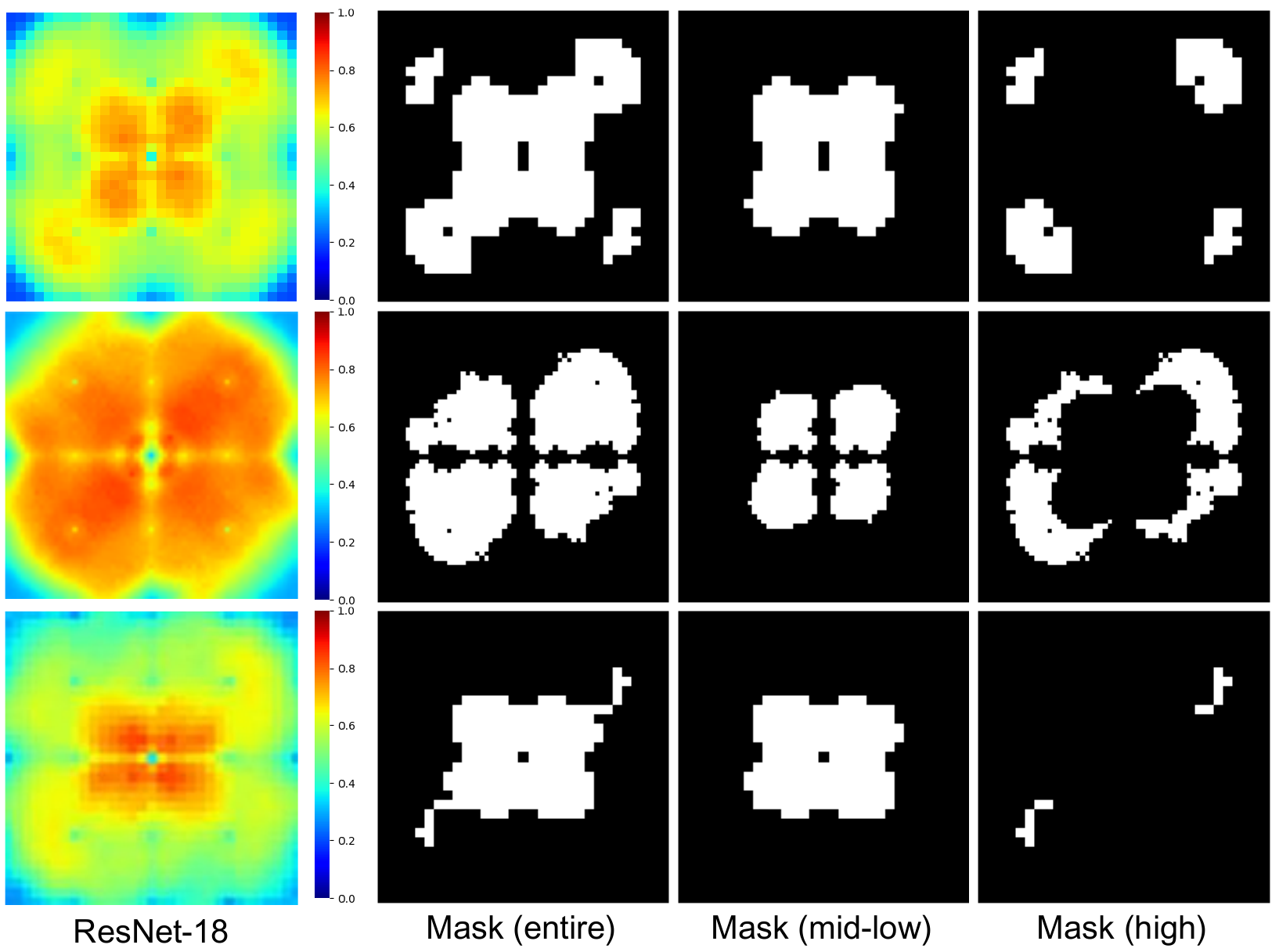}
\caption{The Fourier heat maps and different masks for ResNet-18 evaluated on different datasets. The first column provides the Fourier heat maps on CIFAR-10, CIFAR-100, and GTSRB (from top to bottom) respectively; the second column shows the distribution of all sensitive frequencies; the third column shows the distribution of mid-low sensitive frequencies; the last column shows the the distribution of high sensitive frequencies. }
\end{center}
\end{figure}

\subsection{Comparisons with Previous Methods}
One of the key contributions of this paper is that we propose a new method to generate the trigger samples. It is necessary to compare the visual quality of the trigger samples generated by different methods. To this purpose, we compare the proposed method with three most representative methods, i.e., unrelated samples (URS), logo based trigger (LBT), noise based trigger (NBT). In detail, URS uses images that are not related to the original task of the host model as the triggers \cite{adi:2018, Zhang:AsiaCCS}. LBT simply adds a logo such as character(s) and specific pattern to a clean image to construct the trigger sample \cite{Zhang:AsiaCCS}. NBT uses a pre-defined noise to play the role of the trigger \cite{Zhang:AsiaCCS, Zhong:2020}.

\begin{table*}[!t]
\centering
\caption{Performance comparison for watermarked models using different label assignment strategies. ``RSL'' is short for ``randomly selected label'', and ``NL'' is short for ``new label''.}
\label{tab:my-table}
\begin{tabular}{c|c|cc|cc|cc|cc|cc}
\hline\hline
\multirow{2}{*}{Dataset}  & Model & \multicolumn{2}{c|}{VGG-19} & \multicolumn{2}{c|}{ResNet-18} & \multicolumn{2}{c|}{ResNet-56} & \multicolumn{2}{c|}{DenseNet-121} & \multicolumn{2}{c}{WideResNet-34} \\ \cline{2-12}
  & Strategy & \multicolumn{1}{c|}{$\text{Acc}_\text{o}$} & $\text{Acc}_\text{w}$ & \multicolumn{1}{c|}{$\text{Acc}_\text{o}$} &$\text{Acc}_\text{w}$& \multicolumn{1}{c|}{$\text{Acc}_\text{o}$} & $\text{Acc}_\text{w}$ & \multicolumn{1}{c|}{$\text{Acc}_\text{o}$} & $\text{Acc}_\text{w}$ & \multicolumn{1}{c|}{$\text{Acc}_\text{o}$} & $\text{Acc}_\text{w}$ \\ \hline

\multirow{2}{*}{CIFAR-10}  & RSL      & \multicolumn{1}{c|}{90.74\%} & 96\% & \multicolumn{1}{c|}{92.94\%} & 98\% & \multicolumn{1}{c|}{82.33\%}   & 91\%   & \multicolumn{1}{c|}{94.37\%}   & 98\%   & \multicolumn{1}{c|}{94.45\%}   & 96\%   \\ \cline{2-12}
                           & NL       & \multicolumn{1}{c|}{91.32\%} & 97\% & \multicolumn{1}{c|}{93.88\%} & 99\% & \multicolumn{1}{c|}{83.41\%} & 93\% & \multicolumn{1}{c|}{94.48\%} & 98\% & \multicolumn{1}{c|}{94.53\%} & 99\% \\ \hline
\multirow{2}{*}{CIFAR-100} & RSL      & \multicolumn{1}{c|}{64.12\%} & 93\% & \multicolumn{1}{c|}{74.65\%} & 97\% & \multicolumn{1}{c|}{70.25\%}   & 94\%   & \multicolumn{1}{c|}{76.96\%}   & 96\%   & \multicolumn{1}{c|}{77.59\%}   & 96\%   \\ \cline{2-12}
                           & NL       & \multicolumn{1}{c|}{67.34\%} & 95\% & \multicolumn{1}{c|}{75.72\%} & 99\% & \multicolumn{1}{c|}{72.80\%} & 98\% & \multicolumn{1}{c|}{78.43\%} & 97\% & \multicolumn{1}{c|}{78.71\%} & 98\% \\ \hline
\multirow{2}{*}{GTRSB}     & RSL      & \multicolumn{1}{c|}{97.06\%} & 97\% & \multicolumn{1}{c|}{96.72\%} & 99\% & \multicolumn{1}{c|}{97.49\%}   & 95\%   & \multicolumn{1}{c|}{97.35\%}   & 96\%   & \multicolumn{1}{c|}{97.78\%}   & 98\%   \\ \cline{2-12}
                           & NL       & \multicolumn{1}{c|}{97.28\%} & 98\% & \multicolumn{1}{c|}{96.78\%} & 99\% & \multicolumn{1}{c|}{98.23\%} & 97\% & \multicolumn{1}{c|}{97.57\%} & 97\% & \multicolumn{1}{c|}{98.51\%} & 99\% \\ \hline\hline
\end{tabular}
\end{table*}

The above strategies can be all applied to model verification through analyzing the prediction results of the target model given a certain number of trigger samples. However, in terms of imperceptibility, the above strategies are not desirable for application scenarios. Fig. 8 provides some trigger samples obtained by different methods. Although URS may not significantly impair the performance of the host model on its original task, the irrelevant image content may arouse suspicion of the attacker. LBT and NBT introduce noticeable artifacts in the trigger image, which will impair imperceptibility. Moreover, the noticeable artifacts may expose how the trigger signal was constructed, thereby threatening security. As shown in Fig. 8, the visual difference between the clean samples and the trigger samples generated by the proposed method is small, indicating that the proposed method keeps the imperceptibility of trigger signal very well and therefore is very suitable for practice.

Table X and XI provide more quantitative results. Both the original task and the watermarking task are tested. Moreover, mean PSNR and mean SSIM are used to measure the distortion between the trigger samples and their clean versions. We train each model from scratch. Although different trigger samples can be assigned with different labels, it will surely increase the difficulty of the learning for the model since in this case the model needs to learn various mappings that are not related to the original task which will impair the generalization ability of the model. Therefore, for fair comparison, for each method to be compared, all the trigger samples share the same ground-truth label specified by random in advance (and it should not be the label of the original clean sample). Some examples of trigger samples have been shown in Fig. 8. It is inferred that the proposed method not only achieves better visual quality for the trigger samples, but also shows better performance on both the original task and the watermarking task. It can be said that by adding well-designed perturbation in the frequency domain, the distortion between the clean sample and the trigger sample can be kept low. Meanwhile, as the perturbation will be spread throughout the entire spatial domain, it will better facilitate the learning of the trigger signal.

\begin{table}[!t]
\centering
\caption{Performance comparison between different perturbation strategies for VGG-19. PSNR and SSIM use mean values.}
\begin{tabular}{c|c|c|c|c|c}
\hline\hline
Dataset & Strategy & $\text{Acc}_\text{o}$ & $\text{Acc}_\text{w}$ & PSNR & SSIM\\
\hline
\multirow{2}{*}{CIFAR-10} & Mid-low & 91.32\% & 97\% & 38.85 & 0.9884 \\
 & High & 90.76\% & 96\% & 39.81 & 0.9886 \\
\hline
\multirow{2}{*}{CIFAR-100} & Mid-low & 67.34\% & 95\% & 36.62  & 0.9851 \\
 & High & 65.81\% & 93\% & 37.94 & 0.9873 \\
\hline
\multirow{2}{*}{GTRSB} & Mid-low & 97.28\% & 98\% & 36.64  & 0.9838 \\
 & High & 97.22\% & 96\% & 38.58 & 0.9843 \\
\hline\hline
\end{tabular}
\end{table}

\begin{table}[!t]
\centering
\caption{Performance comparison between different perturbation strategies for ResNet-18. PSNR and SSIM use mean values.}
\begin{tabular}{c|c|c|c|c|c}
\hline\hline
Dataset & Strategy & $\text{Acc}_\text{o}$ & $\text{Acc}_\text{w}$ & PSNR & SSIM\\
\hline
\multirow{2}{*}{CIFAR-10} & Mid-low & 93.88\% & 99\% & 36.14 & 0.9862 \\
 & High &92.72\% & 97\% & 39.27 & 0.9885 \\
\hline
\multirow{2}{*}{CIFAR-100} & Mid-low & 75.72\% & 99\% & 37.46  & 0.9927 \\
 & High & 74.29\% & 96\% & 38.63 & 0.9954 \\
\hline
\multirow{2}{*}{GTRSB} & Mid-low & 96.78\% & 99\% & 38.53  & 0.9935 \\
 & High &96.63\% & 98\% & 40.84  & 0.9971 \\
\hline\hline
\end{tabular}
\end{table}

\subsection{Ablation Study}
In the proposed method, the label assigned to all the trigger samples can be either a randomly selected label or a new label. The randomly selected label should not be the one belonging to the original clean sample. We conduct experiments to evaluate the impact of using different label assignment strategies on both the original task and the watermarking task. Except for the assigned label, all the other experimental settings are same as each other. Table XII provides the experimental results from which we can find that using a new label is superior to using a randomly selected label since the classification accuracies on the two tasks for the former are significantly higher than that for the latter. The reason is that, using a randomly selected label may inevitably distort the decision boundary of the model on the original task, while using a new label maps the trigger samples to a new category so that the learning of the original task and the learning of the watermarking task can be separated to a certain extent, thereby allowing both the original task and the watermarking task to be better performed.

We further analyze the impact caused by perturbing different frequency areas. Taking ResNet-18 for explanation, Fig. 9 shows the Fourier heat maps and different masks evaluated on different datasets. It can be found that different frequency areas have different sensitive-frequency distributions. It is necessary to analyze their impact on the performance of the model. Table XIII and XIV have given the experimental results, from which we can find that the mean PSNRs and the mean SSIMs of the trigger samples due to high frequency perturbation are higher than that due to mid-low frequency perturbation. The reason is that mid-low frequency perturbation results in more distortion in the spatial domain while high frequency perturbation generally distorts local details of an image. However, from the viewpoint of functionality, it can be observed that using mid-low frequency perturbation achieves better performance on the original task and the watermarking task. It is due to the reason that mid-low frequency perturbation can better facilitate model learning, which has been analyzed in the previous section.

\section{Conclusion and Discussion}
Protecting the intellectual property of deep models under the black-box condition against infringement is a very important and challenging problem. How to ensure imperceptibility and robustness of black-box DNN model watermarking is urgently to be solved. Existing methods use trigger samples to achieve black-box model watermarking. Although they can be used for verifying the ownership of the target model, they introduce noticeable visual distortion into the trigger samples. It impairs the imperceptibility of the embedded watermark. Moreover, these methods do not take into account attacks applied to the trigger samples. As a result, the robustness of the watermark is limited. To deal with the above problems, this paper presents a novel method for black-box DNN model watermarking by applying Fourier perturbation analysis and frequency sensitivity clustering. By crafting the trigger samples in the frequency domain, both the original task and the watermarking task can be better performed, which has been verified by our extensive experiments. Additionally, the trigger generation strategy introduced in this paper takes into account the influence caused by the model, which makes the proposed method interpretable.

On the other hand, it should be also admitted that we cannot ensure that the proposed watermarking system is robust against all the real-world attacks because it is surely impossible for us to foresee all the attacks performed by the adversary. Actually, even for the existing methods, they only resist specific attacks. Therefore, in the future, we will further improve the robustness of the proposed method by taking into account more attacks. We will also extend the proposed method to generative models. We hope this
attempt can inspire more advanced works.


\end{document}